\documentclass[style=twocolumn,journal=nalefd,manuscript=communication]{achemso}

\usepackage{epstopdf}
\usepackage[version=3]{mhchem} 
\usepackage[T1]{fontenc}       

\usepackage{listings}

\author{Victor Dmitriev}
\author{Wagner Castro}
\email{wagnerormanes@yahoo.com.br}
\affiliation{Department of Electrical Engineering, Federal University of Para,
	PO Box 8619, Agencia UFPA,  66075-900, Belem,  Para, Brazil.}
\author{Clerisson Nascimento}
\affiliation{Scholarship from CNPq - Brasil}

\title{Dynamically controllable graphene three-port circulator}

\begin{document}
\begin{abstract}
  A  new type of the graphene-based three-port circulator is suggested and analysed. The cross-section of the  component presents a three-layer structure consisting of a layer of silicon,  of silica and  of graphene. In-plane figure resembles a common microwave  nanostrip circulator with a circular graphene resonator and three  waveguides symmetrically connected to it. The graphene  is magnetized normally to its plane by a DC magnetic field. The numerical  simulation  demonstrates the isolation of -15 dB and insertion losses  of -2 dB in 6.98 \% frequency band with the central frequency  8.23THz.
\end{abstract}

\section{Introduction}

Circulators are important passive nonreciprocal components which are used both 
in microwave and optical frequency region. These devices permit mitigate the harmful influence of reflections on the sources of electromagnetic waves. In microwaves, the most popular circulators are based on waveguide and microstrip technology with magnetized ferrite elements \cite{microstripFerrite}.  Optical circulators have been also discussed in literature \cite{optical1, optical5, optical6, optical7}.

In THz region, one of the promising tendencies is graphene technology which allows one, in particular, to construct nonreciprocal devices. Recently, an interesting idea of an edge-guided graphene circulator was suggested in \cite{multGraphCirc}.
The device consists of multi-layer graphene-dielectric waveguides and this can complicate its production technology.

In this paper, we suggest a graphene circulator with a very simple structure 
which can be used in  THz and infrared frequency region. It consists of  graphene nanostrip waveguides connected to a graphene circular resonator. 

\section{Problem description}

\begin{figure}[h!]
	\centerline{
		\includegraphics[scale=1]{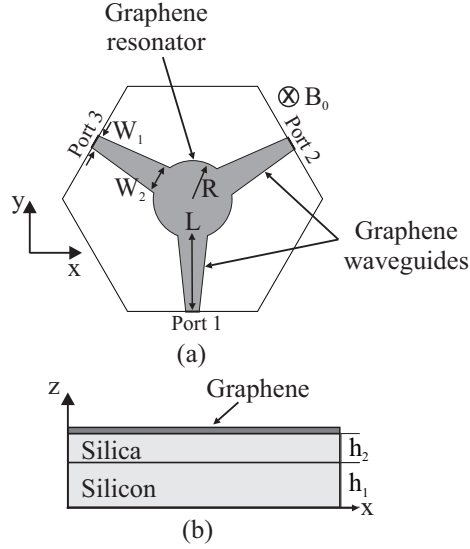}}
	\caption{(Color online) Schematic representation of graphene circulator: a) Top view and b) side view, ${\bf B}_0$ is DC magnetic field.
	}
	\label{figure1}
\end{figure}

The schematic representation of the circulator is shown in Fig. \ref{figure1}a and \ref{figure1}b. The width of the waveguides are  $w_1 = 140$ nm and $w_2 = 200$, that are connected to a circular graphene resonator with the radius  $320$ nm. The waveguides and the resonator are placed on a lossless dielectric substrate of silicon and silica with the thickness  $h_{1}=$ 200 nm and $h_{2}=$ 600 nm, with relative permittivity $\varepsilon_1 = $ 11.9 and  $\varepsilon_1 = $ 2.09,  respectively.

In numerical simulations by the commercial software COMSOL \cite{comsolSite}, we have used the following parameters of the graphene conductivity tensor \cite{grafEquation}:

\begin{eqnarray}
\label{eq:sig1}
\sigma_{xx} = \frac{2D}{\pi} \frac{1/\tau - i\omega}{ \omega_{c}^{2} - ( \omega + i / \tau ) ^2}, \\
\label{eq:sig2}
\sigma_{xy} = - \frac{2D}{\pi} \frac{\omega_c}{ \omega_{c}^{2} - ( \omega + i / \tau ) ^2}.
\end{eqnarray}
where $D = 2 \sigma_0 \epsilon _F / \hbar$ is the drude weight, $\sigma_0$ is the miminum conductivity of graphene, $\epsilon_F$ is related to chemical potential of graphene $\omega_c = {eBv_{F}^{2}}/{\epsilon_F}$ is the cyclotron frequency, $\hbar$ is the reduced Planck's constant, $e$ is the electron charge, $\tau$ is the relaxation time, $\omega$ is the frequency of the incident wave and $i=\sqrt{-1}$. 

For numerical proposition, graphene is modeled as a bulk material with a bulk conductivity tensor given by $[\sigma_v] = [\sigma_s]/\Delta$, being $[\sigma_s]$ the surface conductiviry tensor, which its component are given by (\ref{eq:sig1})-(\ref{eq:sig2}), and $\Delta$ is the thickness of the graphene. In all calculations $\Delta = 1$ nm.

\begin{figure}[h!]
	\centerline{
		\includegraphics[scale=1]{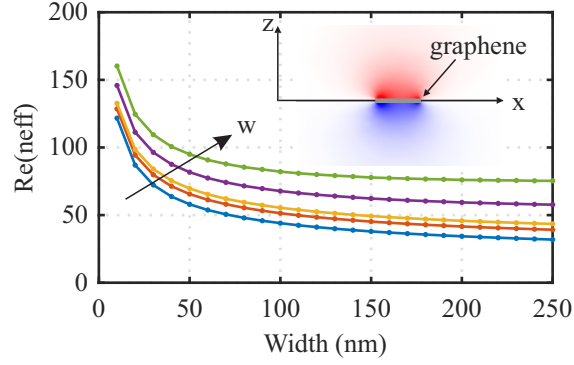}}
	\caption{Width dependency of effective mode index for (upwards) 5 THz, 8 THz, 10 THz, 15 THz, 20 THz. Insert is the $E_z$  profile of guided mode in graphene strip.}
	\label{fig:neff}
\end{figure}

It is well known that graphene stips can suport two kinds of guided surface plasmon polaritons (SPP) modes, namelly a guide mode and edge mode, which are discussed in details in \cite{Nikitin2011, HE2013, Sheng2015}. In our case, we have chosen the former one, once the eletromagnetic field in that mode is concentrated in the center of the ribbon, as shown in the insert in Fig \ref{fig:neff}.  The dependence of effective refractive index for different values of frequency, with respect to width of the geraphene ribbon is shown in Fig. \ref{fig:neff}.
 
In the following, the transmission coefficient $S_{ij} = P_j/P_i$ of a signal injected in port $j$ and received by port $i$ was calculated. The percentage bandwidth of circulator is defined as $BW = (f_2 - f_1)/f_0 \times 100$, where $f_2$ and $f_1$  are the limit frequency value so that the transmission level on a transmission port remain above - 2 dB and the isolation level on the isolation port remain below - 15 dB at the same time. $f_0$ is the central frequency of operation.

\section{Numerical Result}

\begin{figure}[h!]
	\centerline{
		\includegraphics[scale=1]{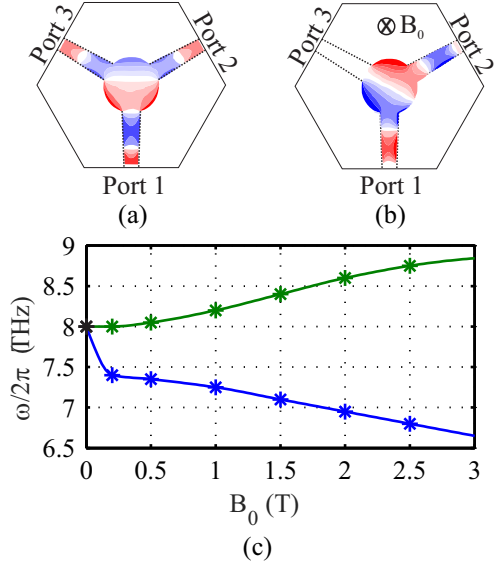}}
	\caption{(Color online) $E_z$ field profile for a) nonmagnetized and b) magnetized case, and c) magnetic field dependency of rotating modes.
	}
	\label{fig:desdob}
\end{figure}

The working principle of the circulator is similar to that of the microstrip one. The guided SPP wave in the input waveguide excites in the circular resonator in the non-magnetic state two degenerate clockwise $\omega^{+}$ and anticlockwise $\omega^{-}$ rotating modes. These two modes produce a dipole aligned to the input waveguide, which leads to a transmission in both output ports, as shown in Fig. \ref{fig:desdob}a. A DC magnetic bias of $B_0 =1.1$ T brakes the degenerency and makes the field pattern of the standing wave to rotate by 60$^{\circ}$ aligning thus the node of dipole with the isolation of port 3 as shown in Fig. \ref{fig:desdob}b. The dependency of these two rotating dipole modes with respect to magnetic field is shown in Fig. \ref{fig:desdob}c.

\begin{figure}[h!]
	\centerline{
		\includegraphics[scale=1]{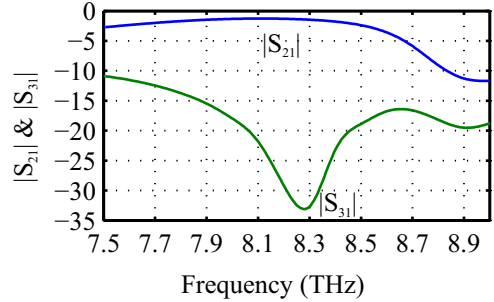}}
	\caption{(Color online) Frequency response of circulator.
	}
	\label{fig:S21S31}
\end{figure}

The calculated frequency characteristics are shown in Fig. \ref{fig:S21S31}.  At a central frequency of 8.23 THz, the device presents a transmission coefficient of -1.2 dB, a isolation of -33 dB, with a bandwidth of 6.98 \% in a transmission level of -2 dB and a isolation level of -15 dB.

\subsection{Control by chemical potential of graphene}
\begin{figure}[h!]
	\centerline{
		\includegraphics[scale=1]{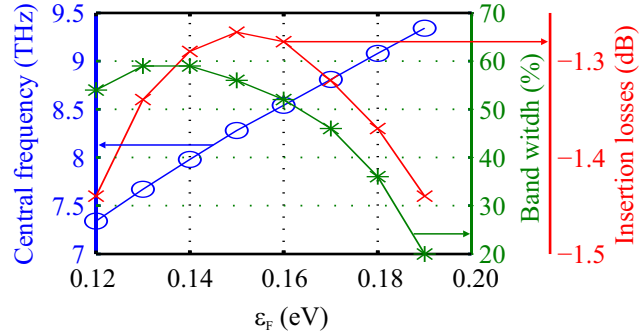}}
	\caption{(Color online) Chemical potential dependency of the central frequency (circle marks), bandwidth (asterisk marks) and of the insertion losses (x marks).
	}
	\label{fig:mucdep}
\end{figure}

By changing chemical potential of graphene, one can change the carrier density in the ribbon, and consequently, the values of the conductivity tensor components. This leads to a possibility of dynamical control of  the circulator responses. One can see from, Fig. \ref{fig:mucdep}, that the central frequency of circulator operation shifts to higher frequencies, when the chemical potencial increases. The opposite situation occurs by diminishing this value. Besides by changing $\mu_c$, the values of the insertion losses and the bandwidth get worst than those for $\mu_c = 0.15$ eV. It can be explained by the fact that the stationary dipole is not property aligned to the output port, but one can see that for values between 0.12 eV and 0.20 eV, these levels are acceptable, as shown in the same figure. 

\section{Conclusions}
In this work we have suggested and confirmed by  numerical simulations a possibility of  a controllable three-port circulator graphene-based circulator. This component with a very simple structure can be used in THz and infrared circuits.

\begin{acknowledgement}
This work was supported by Brazilian agency CNPq.
\end{acknowledgement}



\end{document}